\documentclass[article,nofootinbib,preprintnumbers,floats,aps]{revtex4}

\usepackage{amsmath}
\usepackage{amssymb}
\usepackage{graphicx}
\usepackage{epsfig}
\usepackage{bm}
\usepackage{units}
\usepackage{subfigure}
\usepackage{hyperref}
\usepackage{breakurl}

\newcommand{\be}{\begin{equation}}
\newcommand{\ee}{\end{equation}}
\newcommand{\bea}{\begin{eqnarray}}
\newcommand{\eea}{\end{eqnarray}}
\newcommand{\ben}{\begin{enumerate}}
\newcommand{\een}{\end{enumerate}}
\newcommand{\bit}{\begin{itemize}}
\newcommand{\eit}{\end{itemize}}

\newcommand{\vv}{\boldsymbol}					

\newcommand{\bert}{\raise-0.45mm\hbox{\Large$\Box$}}

\begin{document}

\preprint{MIT-CTP-4060}
\preprint{0908.3190 [physics.flu-dyn]}

\title{Sprinkler Head Revisited: \\
Momentum, Forces, and Flows in Machian Propulsion}
 
\author{Alejandro Jenkins}\email{ajv@mit.edu} \altaffiliation{Current address: High Energy Physics, 505 Keen Building, Florida State University, Tallahassee, FL 32306-4350, USA; \url{jenkins@hep.fsu.edu}}

\affiliation{Center for Theoretical Physics,
Laboratory for Nuclear Science and Department of Physics,
Massachusetts Institute of Technology, Cambridge, MA 02139, USA}

\date{Aug. 2009, last revised Jun. 2011; to appear in Eur. J. Phys. {\bf 32}} 

\begin{abstract}

Many experimenters, starting with Ernst Mach in 1883, have reported that if a device alternately sucks in and then expels a surrounding fluid, it moves in the same direction as if it only expelled fluid.  This surprising phenomenon, which we call {\it Machian propulsion}, is explained by conservation of momentum: the outflow efficiently transfers momentum away from the device and into the surrounding medium, while the inflow can do so only by viscous diffusion.  However, many previous theoretical discussions have focused instead on the difference in the shapes of the outflow and the inflow.  Whereas the argument based on conservation is straightforward and complete, the analysis of the shapes of the flows is more subtle and requires conservation in the first place.  Our discussion covers three devices that have usually been treated separately: the reverse sprinkler (also called the inverse, or Feynman sprinkler), the putt-putt boat, and the aspirating cantilever.  We then briefly mention some applications of Machian propulsion, ranging from microengineering to astrophysics. \\



\end{abstract}

\maketitle

\section{Introduction}
\label{sec:intro}

\begin{figure} [b]
\begin{center}
	\includegraphics[width=0.3\textwidth]{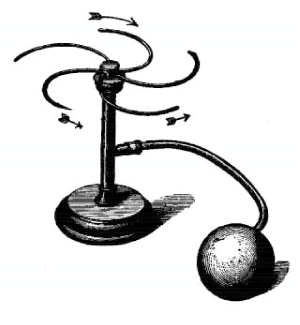}
\end{center}
\caption{\small The reaction wheel, as shown in \href{http://echo.mpiwg-berlin.mpg.de/ECHOdocuViewfull?pn=290&url=/mpiwg/online/permanent/einstein_exhibition/sources/Q179XRYG/pageimg&viewMode=images&tocMode=thumbs&tocPN=1&searchPN=1&mode=imagepath&characterNormalization=reg}{Figure 153a} of Ernst Mach's {\it Mechanik}  \cite{Mach}.  The two short arrows show the direction of the air expelled when the rubber ball is squeezed.  The long arrow shows the direction in which the device turns.  (Image in the public domain.)\label{fig:reactionwheel}}
\end{figure}

In chapter III, section III of his 1883 text on mechanics, physicist and philosopher of science Ernst Mach analyzes the behavior of a ``reaction wheel,'' illustrated in Fig. \ref{fig:reactionwheel}, which alternately expels and then sucks in air as the experimenter squeezes and then releases the hollow rubber ball.  He observed that ``the wheel [continues] to revolve rapidly in the same direction as it did in the case in which we blew into it.''  In other words, the effect of sucking a given volume of air does not cancel the effect of blowing it out.  Indeed, Mach did not notice {\it any} motion of the reaction wheel when it was made to suck in air \cite{Mach}.  Similar observations were later made independently by others (e.g.,  \cite{Kirkpatrick,Belson}).

Mach's reaction wheel is a close analog of the so-called reverse (or inverse) sprinkler problem, made famous by theoretical physicist (and Nobel laureate) Richard P. Feynman's bestselling book of  personal reminiscences, published in 1985 \cite{Feynman-anecdote}.  As a graduate student at Princeton University in the early 1940's, Feynman attempted to determine which way a sprinkler would turn if it were submerged and made to suck in the surrounding water.  His improvised experiment in the rooms of the university's  cyclotron laboratory ended with the explosion of a large glass bottle filled with water.  (For other first-hand accounts of this incident, see \cite{Wheeler,Creutz}.)

Various experimental and theoretical analyses of the reverse sprinkler have established that it turns in the direction opposite to that of the regular sprinkler, but far more weakly. In fact, were it not for the viscosity of the fluid, the reverse sprinkler would experience only a transient torque as the flow commences, and no torque at all in the steady state.  This can be understood by invoking the conservation of (angular) momentum, as we shall review in Sec. \ref{sec:momentum}.

\begin{figure} [t]
\begin{center}
	\includegraphics[width=0.5\textwidth]{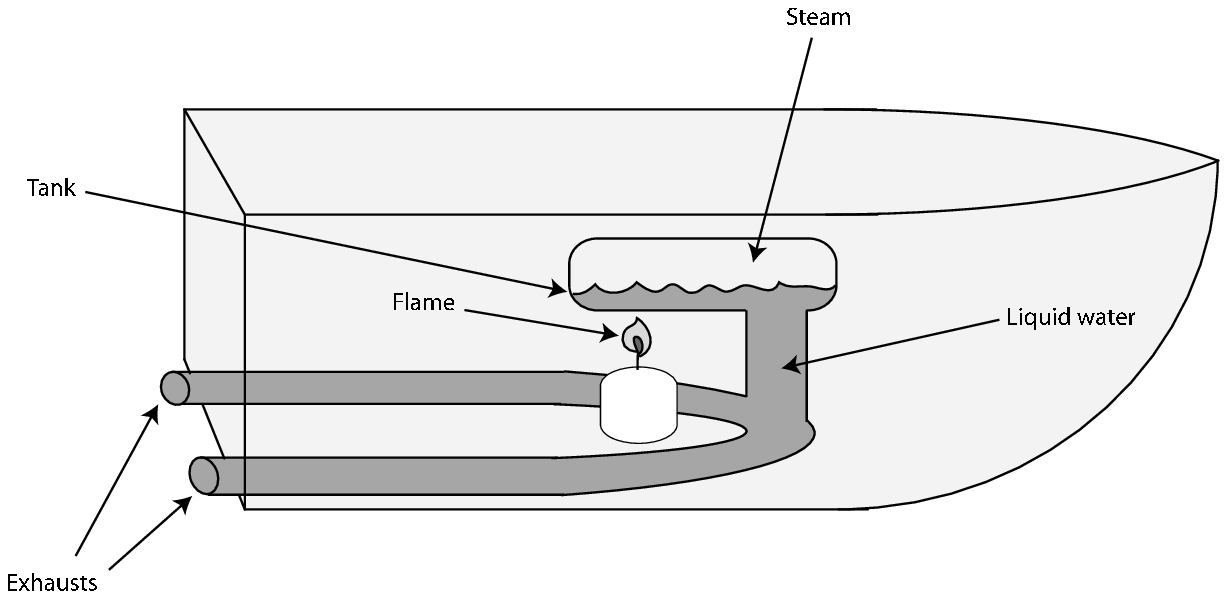}
\end{center}
\caption{\small Schematic illustration of the putt-putt boat, a toy propelled by an oscillation of the steam pressure inside the tank that causes water to be alternately expelled from the submerged exhausts and then drawn back in. \label{fig:putt-putt}}
\end{figure}

Another striking manifestation of the same underlying physics is the so-called putt-putt (or pop-pop) boat, a toy boat powered by a candle that heats an internal tank filled with water and connected to submerged exhausts (see Fig. \ref{fig:putt-putt}).  As the heat of the candle causes water to evaporate, the pressure of the steam pushes liquid out of the tank.  When the toy is working properly, the heat of the flame is not so intense as to drive the liquid out of the tank completely. Instead, some of the steam quickly recondenses on the relatively cool tank walls, causing the pressure in the tank to drop and water to be drawn in through the exhausts.  As the water level rises and less surface area on the tank walls is available for condensation, the pressure of the steam increases again.  This leads to a cycle that propels the boat forward and causes a noisy vibration that gives the toy its name. \cite{putt-putt,Finnie}

\begin{figure} [t]
\begin{center}
	\includegraphics[width=0.6\textwidth]{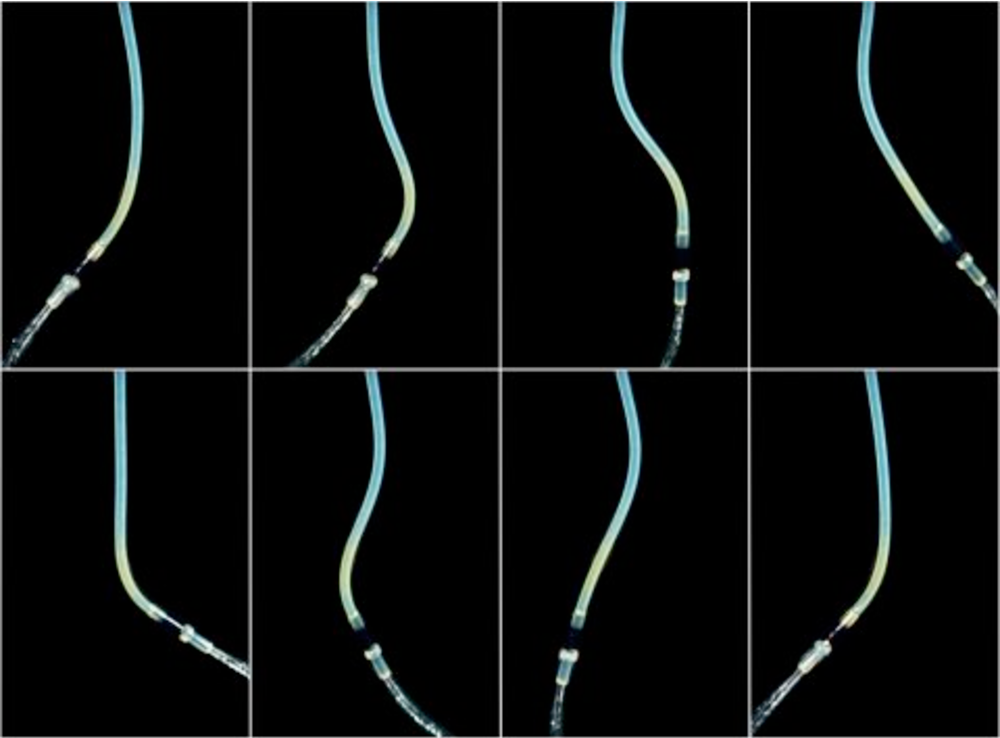}
\end{center}
\caption{\small Time-lapse pictures illustrating the garden-hose instability.  Images by Olivier Doar\'e (ENSTA) and Emmanuel de Langre (\'Ecole Polytechnique) \cite{deLangre1,deLangre2}, used here with the permission of the authors.\label{fig:gardenhose}}
\end{figure}

A third instance of this same effect is observed in the behavior of cantilevered pipes.  When such a pipe expels fluid, it is subject to a ``garden-hose instability'' which can lead to uncontrolled oscillations \cite{AxialFlow}, caused by the misalignment of the momentum of the outgoing fluid with the axis of the pipe (think of an unsupported garden hose running at full blast, as pictured in Fig. \ref{fig:gardenhose}).  This instability is far less severe in the case of aspirating pipes, for the same reason that, for equivalent rates of flow, the torque on the reverse sprinkler is much smaller than on the regular sprinkler. \cite{Paidoussis}

Since, to our knowledge, Mach was the first to describe it clearly in print, we will use the term ``Machian propulsion'' to refer to the fact that a device that alternately aspirates and then discharges fluid moves in the same direction as a device that only discharges.  More generally, we will use the same term to refer to the smallness of the force (or torque) on the aspirating device, relative to the force (or torque) on the device that discharges fluid at the same rate.

\begin{figure}[b]
\centering
\subfigure[]{\includegraphics[width=0.36 \textwidth]{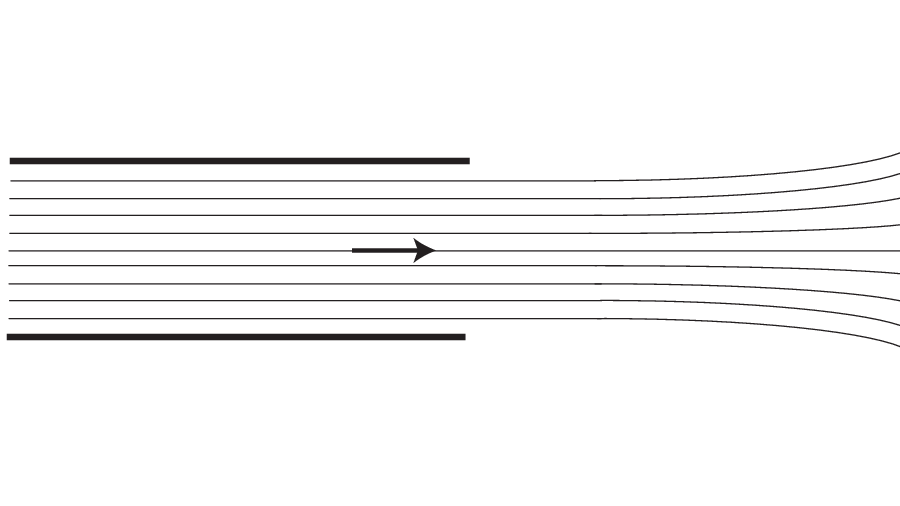}} \hskip 3.2 cm
\subfigure[]{\includegraphics[width=0.31 \textwidth]{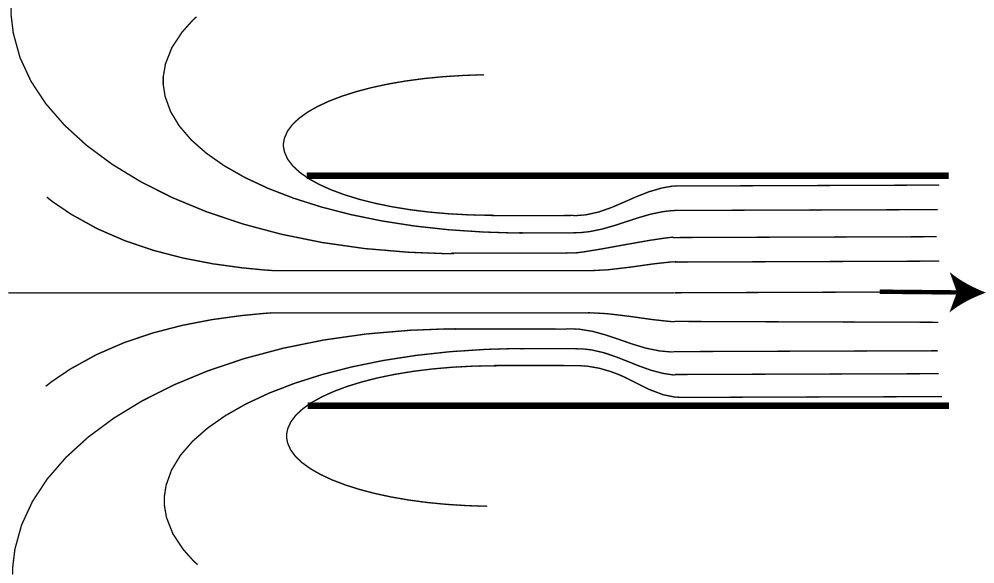}}
\caption{Streamlines for the flow (a) expelled from the mouth of a tube, and (b) aspirated into the mouth of the tube. \label{fig:flows}}
\end{figure}

Many theoretical treatments of these systems, starting with Mach's, have emphasized that the fluid expelled forms a jet, whereas the flow sucked in ``comes in from all directions,'' as shown schematically on Fig. \ref{fig:flows}.  This difference in the shape of the flows has often been adduced as an explanation of Machian propulsion, but, even though such an argument does establish that there is no time-reversal symmetry between aspiration and discharge, by itself it does not take us very far towards understanding the forces involved during each phase.  In fact, we shall see that one should start from conservation of momentum in order to understand the shapes of the flows.  Also, the asymmetry of the shapes of the in- and the outflow is due to viscosity, whereas Machian propulsion would be observed in inviscid flow.

\section{Conservation of momentum}
\label{sec:momentum}

Conservation of momentum provides the simplest and most reliable theoretical tool for understanding Machian propulsion.  In the context of the reverse sprinkler problem, this argument was first made clearly in print in 1987, in a brief letter by Alton K. Schultz (a geophysicist) \cite{Schultz}.  Schultz's argument is as follows: as water flows out of a regular sprinkler, it carries away with it an ever increasing quantity of angular momentum about the sprinkler's pivot.  If the sprinkler operated in empty space, this angular momentum would be carried by the water expelled, as it moves away to infinity.  For a sprinkler operating on the earth, this angular momentum is transferred from the water to the earth as the water hits the ground around the sprinkler.

By conservation of angular momentum, the sprinkler must therefore acquire an opposite angular momentum about its pivot.  If the flow is steady, the water's angular momentum increases at a constant rate, and so the sprinkler must experience a constant torque in the opposite direction, which causes it to undergo angular acceleration (until friction and air resistance balance that torque and the sprinkler stops accelerating).

The situation with the reverse sprinkler is very different.  Initially, the water in the tank is still and carries no angular momentum about the sprinkler's pivot.  As the pump is turned on and the flow of water is established, the water in the tank begins to acquire angular momentum, and the sprinkler must therefore experience a corresponding torque in the opposite direction, which makes it accelerate towards the incoming fluid.  The water that is sucked into the reverse sprinkler does not end up carrying away with it any angular momentum: it transfers its angular momentum back to the sprinkler and leaves the tank without any angular momentum.  In other words, in the reverse sprinkler's steady state, the total amount of angular momentum in the water is {\it not} growing: it is a constant quantity, and therefore the reverse sprinkler experiences no torque in the steady state.  When the flow of water stops, the sprinkler experiences a torque in the opposite direction to before, as it gives up its angular momentum and comes to rest.

Thus, if the reverse sprinkler moves without friction or resistance in an ideal fluid, it first accelerates towards the incoming water, then turns with constant angular velocity in its steady state, and finally comes to a stop when the flow of water is shut off.  One complication which was not considered by Schultz but which is discussed in \cite{Jenkins}, is that, since the water has some viscosity, not all of its steady-state angular momentum will be transferred to the sprinkler as the water leaves the tank.  There will be some amount of water flow that does not enter the sprinkler head.\footnote{It is well known that viscosity in a Newtonian fluid is directly related to the diffusion of momentum.}  The corresponding angular momentum will be transferred to the surrounding tank, and, with respect to the tank's frame of reference, the reverse sprinkler will experience a small torque even in the steady state.  This torque tends to make the reverse sprinkler turn towards the incoming water, in the direction opposite to the rotation of the regular sprinkler.\footnote{The fact that the reverse sprinkler experiences no torque for steady, inviscid flow may also be interpreted as a variation on a theorem, the so-called d'Alembert's paradox, which establishes that steady, inviscid flow cannot exert any drag on a solid object \cite{dAlembert} (see also \cite{Batchelor-dAlembert}).  Clearly, such a result must follow from conservation as long as there is no mechanism to diffuse momentum out of a surface enclosing the solid object and the region of steady flow around it, as explained in \cite{Milne-Thomson-dAlembert}.}

All of these predictions are supported by experiment \cite{UMD-web}, though it might be desirable to rigorously test the dependence of the steady-state torque on the viscosity of the fluid.  Wolfgang Rueckner (Harvard) has suggested operating a reverse sprinkler in a fluid whose viscosity depends strongly on temperature ---such as argon gas or liquid glycerin--- and reported some encouraging (but very preliminary) results with the former: while maintaining a constant flow rate, an increase in viscosity seems to result in an increase in the terminal angular velocity of the reverse sprinkler. \cite{Rueckner}  

In short, both the regular and the inverse sprinkler can turn {\it only to the extent that angular momentum may be transferred to the surrounding environment}.  It is straightforward to generalize this argument to the putt-putt boat by considering the conservation of linear, rather than angular, momentum.  Remarkably, this elementary observation suffices to clarify a number of confusions that persist in the scientific literature on Machian propulsion, as we shall see in Secs. \ref{sec:shapes} and \ref{sec:applications}.  

\section{Forces and flow shapes}
\label{sec:forces}

The argument based on momentum conservation is correct and complete, but published treatments have usually focused on finding the forces (or torques) acting on the device in question (for earlier work on this subject, which shows a gradual evolution in the understanding of the reverse sprinkler, see \cite{Forrester,Harvard,UMD1,UMD2,depressurization}.)  As explained in Sec. II of \cite{Jenkins} and then cleverly demonstrated experimentally in \cite{Mungan}, for the reverse sprinkler the relevant torques are produced by a  ``pressure difference effect,'' which imparts to the sprinkler an angular momentum opposite to that of the incoming fluid, and a ``momentum transfer effect,'' by which the aspirated fluid transfers its angular momentum to the sprinkler when it impinges on the tube's inner wall.  In the absence of viscosity, these torques cancel each other out exactly in the steady state, as required by the conservation argument given in Sec. \ref{sec:momentum}.

As a reader (Lewis H. Mammel, Jr.) pointed out to us after \cite{Jenkins} appeared in print, the derivation presented there of the magnitude of the ``momentum transfer'' effect does not reflect the fact that the cross-section of a fluid flow is not constant along a pressure gradient \cite{Mammel}.  For instance, it is clear that an incompressible fluid cannot maintain a constant cross-section $A$ if its speed $v$ is increasing along the direction of the flow, since continuity requires
\be
A_1 v_1 = A_2 v_2~,
\label{eq:continuity}
\ee
and, by Bernoulli's theorem, for an ideal fluid with density $\rho$,
\be
A_2 = \frac{A_1 v_1}{\sqrt{v_1^2 + 2 (P_1 - P_2) / \rho}}~.
\ee
The conservation argument described in Sec. \ref{sec:momentum} suffices to establish that the magnitude of the ``momentum transfer'' effect that is derived in \cite{Jenkins} is correct, but the simplified treatment of the forces that was presented there does not account for the fact that the shape of the flow can, in practice, be quite complicated.\footnote{Some comments on the issue of the shape of the flows were added to the version of \cite{Jenkins} that appears as chapter 6 of \cite{Thesis}.}  This seems an important issue to clarify, because many discussions of Machian propulsion have offered the difference in the forms of the in- and outflows as the fundamental explanation of the phenomenon.

Mach, for instance, claims in \cite{Mach} that the behavior of the reaction wheel of Fig. \ref{fig:reactionwheel} ``results partly from the difference in the motion which the air outside the tube assumes in the two cases.  In blowing, the air flows out in jets [\ldots] In sucking, the air comes in from all sides, and has no distinct rotation.''  James Gleick, in his bestselling biography of Feynman, also explains the behavior of the reverse sprinkler in terms of different shapes on the in- and outflows \cite{Gleick}.  In their treatment of the putt-putt boat, Finnie and Curl explain that toy's propulsion in terms of the shapes of the flows, but also admit that it is possible to deform the flows by placing a nozzle in the mouth of the exhausts, which makes their argument somewhat obscure \cite{Finnie}.  The authors of \cite{scallop} claim that the fact that  ``a candle can be put out by blowing, but not by sucking'' (which reflects the omni-directionality of the inflow, compared to the directionality of the outflow) is equivalent to the phenomenon of Machian propulsion.

The distinct shapes of the out- and inflows, shown in Fig. \ref{fig:flows}, do illustrate the fact that aspiration is different from the time-reversed picture of discharge.  This observation, however, is not enough to explain just how Machian propulsion works, as we will explain in Sec. \ref{sec:shapes}.

\section{Review: Efflux coefficients}
\label{sec:efflux}

In section 40-3 of his {\it Lectures on Physics}, Feynman calls the derivation of the {\it efflux coefficient} (known also as the ``coefficient of discharge'') for a re-entrant discharge tube, ``most beautiful,'' but then gives us ``just a hint'' of how the argument goes \cite{Feynman-lectures}.  Here we will fill in the details.\footnote{The argument presented in this section was first made in \cite{Borda}.  Textbook treatments similar to this derivation include \cite{Milne-efflux} and \cite{Daugherty-efflux}.  As far as possible, we shall avoid both the jargon and the vector calculus notation of advanced fluid mechanics textbooks since they will not be useful in our discussion, for reasons that will become clear.}

\begin{figure} [t]
\begin{center}
	\includegraphics[width=0.35\textwidth]{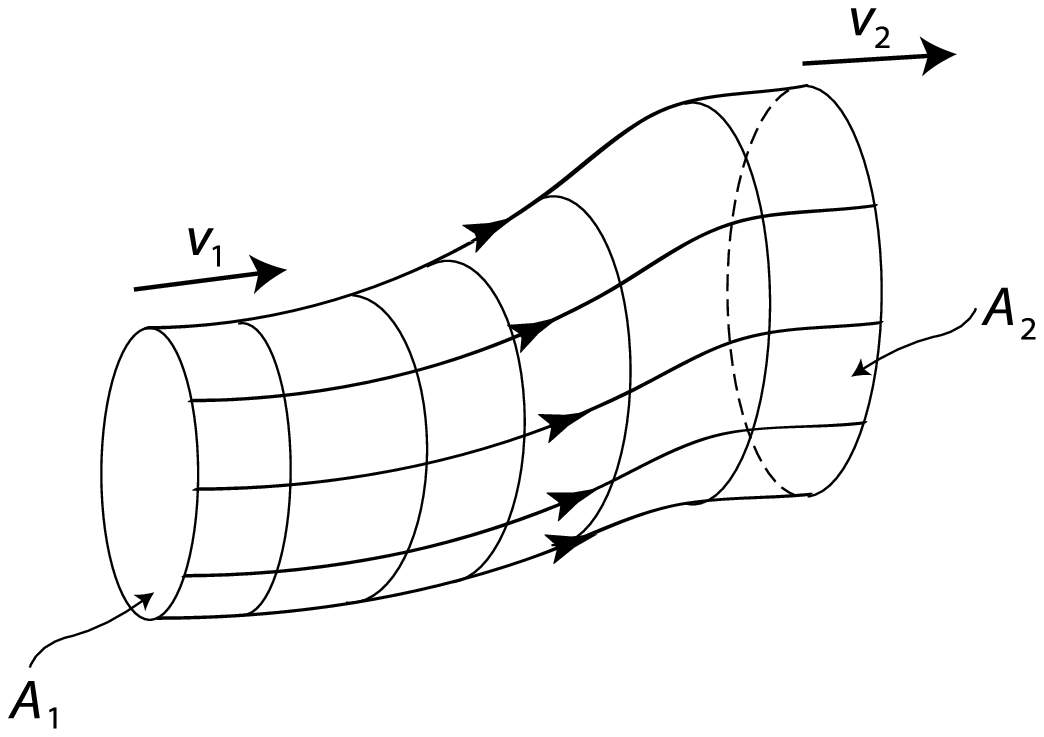}
\end{center}
\caption{\small Steady, irrotational fluid motion along a flow tube, defined by a set of adjacent streamlines. \label{fig:flowtube}}
\end{figure}

Consider the steady motion along a flow tube, defined by a set of adjacent streamlines, as shown in Fig. \ref{fig:flowtube}.  For an irrotational fluid with constant density $\rho$, we see that ---per unit time--- a momentum $\rho v_2^2 \vv A_2$ flows out while a momentum $\rho v_1^2 \vv A_1$ flows in.\footnote{To simplify our notation, we have defined the area vectors $\vv A_{1,2}$ to point along the direction of the corresponding fluid velocities, $\vv v_{1,2}$, with a magnitude equal to the corresponding cross-section, as illustrated in Fig. \ref{fig:flowtube}.}  Therefore, the net force pushing the fluid along this flow tube is
\be
\vv F = \rho (v_2^2 \vv A_2 - v_1^2 \vv A_1)~.
\label{eq:momentum-flow}
\ee 
This is a special case of the ``momentum theorem'' first derived by Euler \cite{Euler}.  For modern, general discussions of this theorem, see \cite{Milne-momentum, Batchelor-momentum, Daugherty-momentum, Potter-momentum}.\footnote{As Feynman explains clearly in \cite{Feynman-lectures}, the total momentum within the flow tube may be varying even if the flow is steady.  The reason is that the velocity field $\vv v$ is defined as function of the point in space, but Newton's laws apply to individual mass elements, which move along the flow.  Steady flow means that the velocity field $\vv v(x,y,z)$ is constant in time, but the mass elements of the fluid may be experiencing a net force the pushes them along the flow tube of Fig. \ref{fig:flowtube}.  The rate at which mass flows through an oriented area element $d \vv a$ with fixed position, is $\rho \vv v \cdot d \vv a$.  In general, the net force on the steady flow enclosed by some surface is equal to the integral of $\vv v (\rho \vv v \cdot d\vv a)$ over that surface.}

\begin{figure}[t]
\centering
\subfigure[]{\includegraphics[width=0.35 \textwidth]{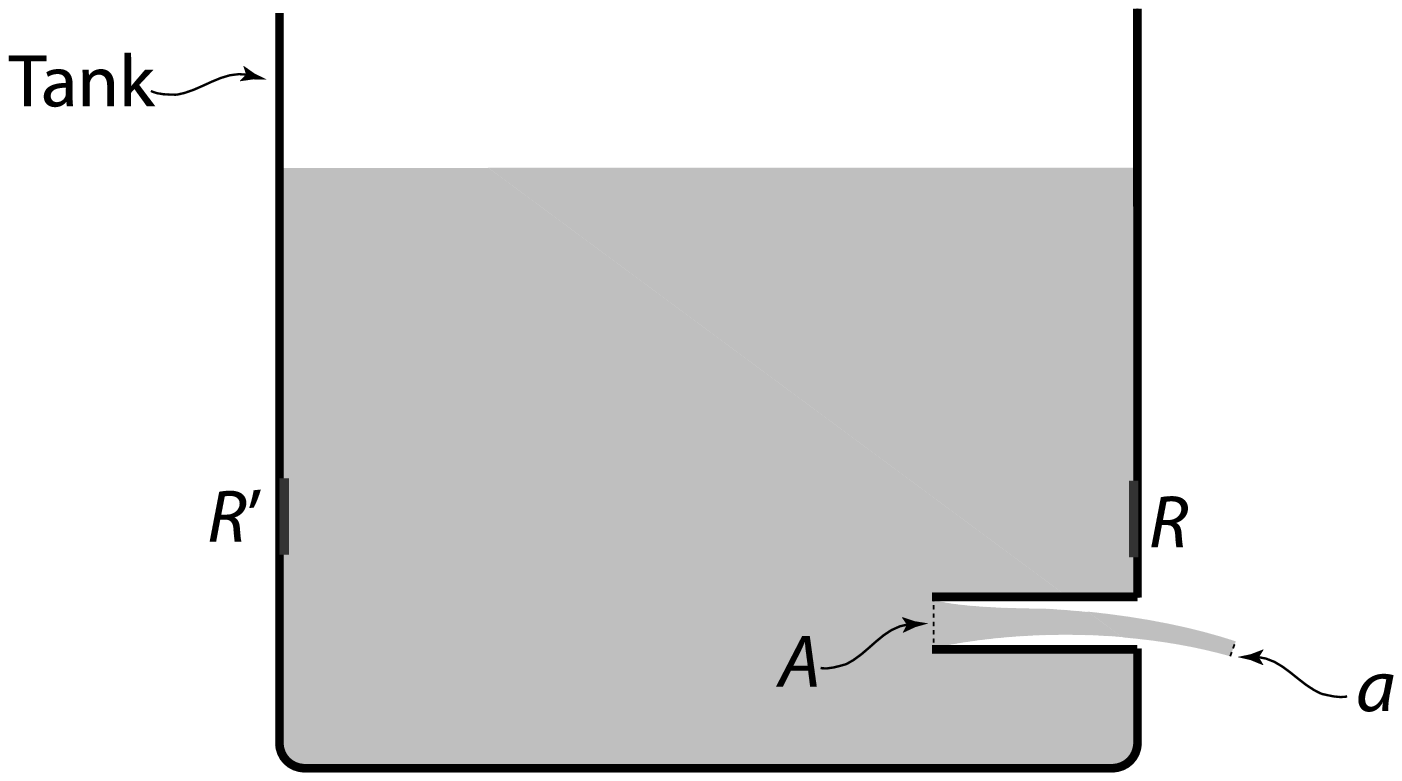}} \hskip 2.5 cm
\subfigure[]{\includegraphics[width=0.32 \textwidth]{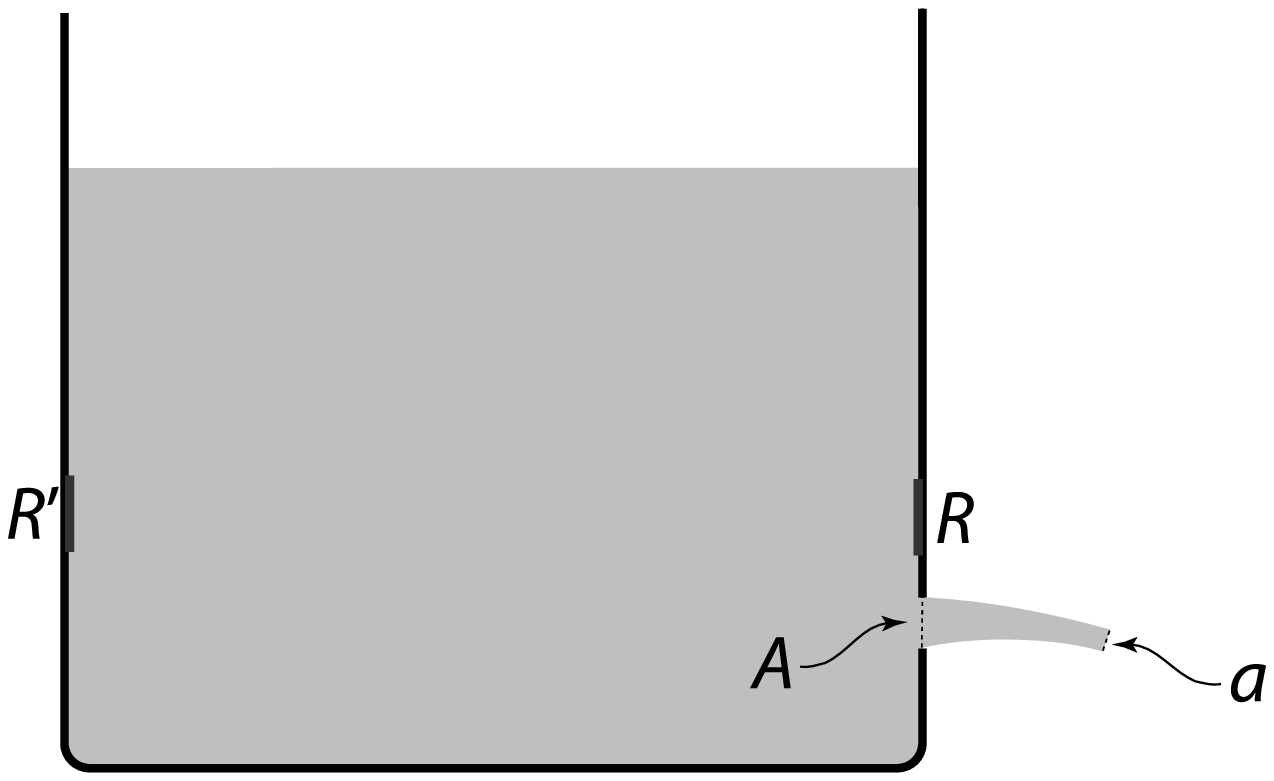}}
\caption{Flows out of a tank, through: (a) a re-entrant discharge tube, and (b) a hole in the tank wall.  The {\it efflux coefficient} is defined as the ratio $a / A$.  In (a), the force exerted on the fluid by the solid wall in region $R$ cancels the force exerted by the opposite region, $R'$.  This is not true in (b), since, for instance, the pressure on $R$ is lower than the pressure on $R'$.
\label{fig:tanks}}
\end{figure}

Consider now a re-entrant discharge tube on a tank, as shown in Fig. \ref{fig:tanks}(a).  This setup has the nice feature that the velocity of the fluid everywhere near the walls of the tank is negligible.  The net horizontal force that accelerates the fluid into the discharge tube must come, originally, from a solid wall pushing on the fluid adjacent to it.  The pushes from opposite sides of the tank wall (e.g., from regions $R$ and $R'$ in Fig. \ref{fig:tanks}(a)) cancel out.  The only net horizontal force therefore comes from the section directly opposite to the mouth of the discharge tube, with area $A$.

Thus, the horizontal force accelerating the fluid is $F = PA$, where $P$ is the hydrostatic pressure on the fluid next to the wall opposite to the mouth of the tube (relative, of course, to the atmospheric pressure outside).  This must also be equal to the rate at which horizontal momentum is pouring out of the tank.  Therefore, if $v$ is the velocity of the jet after the flow has become parallel, we conclude from Eq. (\ref{eq:momentum-flow}) that
\be
F = P A = a \rho v^2~.
\label{eq:reentrant-force}
\ee
Meanwhile, by Bernoulli's theorem,
\be
P = \frac{1}{2} \rho v^2~,
\label{eq:reentrant-Bernoulli}
\ee
which implies that
\be
\frac{a}{A} = \frac{1}{2}~,
\label{eq:reentrant-efflux}
\ee
so that the efflux coefficient in this case is exactly $\nicefrac{1}{2}$.\footnote{Note that, for Eq.  \ref{eq:reentrant-force} to be valid, $a$ must be measured where the streamlines of the jet are parallel and horizontal.  We have neglected both the the outpouring jet's vertical momentum, imparted by gravity, and the resistance from the surrounding air, which gradually slows the jet, causing its cross-section to expand.  The latter effect will be relevant in Sec. \ref{sec:shapes}.}

The efflux coefficient measured for a discharge {\it hole}, as shown in Fig. \ref{fig:tanks}(b), is greater than  $\nicefrac{1}{2}$, because the fluid next to the wall regions above and below the hole is {\it not} at rest and therefore has a lower pressure than the fluid on the opposite side of the tank.  For example, in Fig. \ref{fig:tanks}(b) the push exerted by region $R$ is less than the push exerted by $R'$.  The net horizontal force on the fluid is therefore
\be
F = a \rho v^2 > P A~.
\label{eq:hole-efflux}
\ee
Experimentally, $a/A \simeq 0.62$.\footnote{The theoretical result for an ideal fluid pouring out of a sharp-edged, circular orifice is $\pi / (\pi + 2)$; see \cite{efflux-hole}.}

Advanced textbooks (e.g., \cite{Batchelor-efflux}) sometimes derive Eq. \ref{eq:reentrant-efflux} by integrating the Navier-Stokes equation for inviscid flow (a special case often called the ``Euler equation'').  The Navier-Stokes equation is derived by applying momentum conservation to local fluid elements and cannot be integrated analytically if the flow is viscous (see, e.g., \cite{Feynman-lectures, Navier-Stokes}).  Since in Sec. \ref{sec:shapes} we will need to treat a problem that involves viscosity in a fundamental way, we have not written down a Navier-Stokes equation, but instead worked in terms of global momentum conservation.  In the bargain, we avoid introducing a vector calculus notation which would not have been helpful in making the arguments relevant to the present discussion.  (This perhaps reflects a useful lesson: that not every problem in fluid mechanics is best addressed by the Navier-Stokes equation.)

\section{Shape of inflow}
\label{sec:shapes}

This simple discussion allows us to understand the basic shape of the flows shown schematically in Fig. \ref{fig:flows}.  The outflow forms a jet, with streamlines nearly parallel to the axis of the tube in the region just outside the tube's mouth.  In the case of the inflow, fluid comes in from all sides into the mouth of the tube, then forms a {\it vena contracta} (Latin for ``contracted vein''), where the cross section of the flow is smallest.  The same argument used for the re-entrant discharge tube of the tank now serves to show that, for an ideal fluid, the cross-section of the {\it vena contracta} is half of the cross section of the tube for an ideal fluid.  This behavior was first described by Jean-Charles de Borda in 1766; the re-entrant discharge tube is therefore also called a ``Borda mouthpiece''  \cite{Borda}.

\begin{figure} [t]
\begin{center}
	\includegraphics[width=0.5\textwidth]{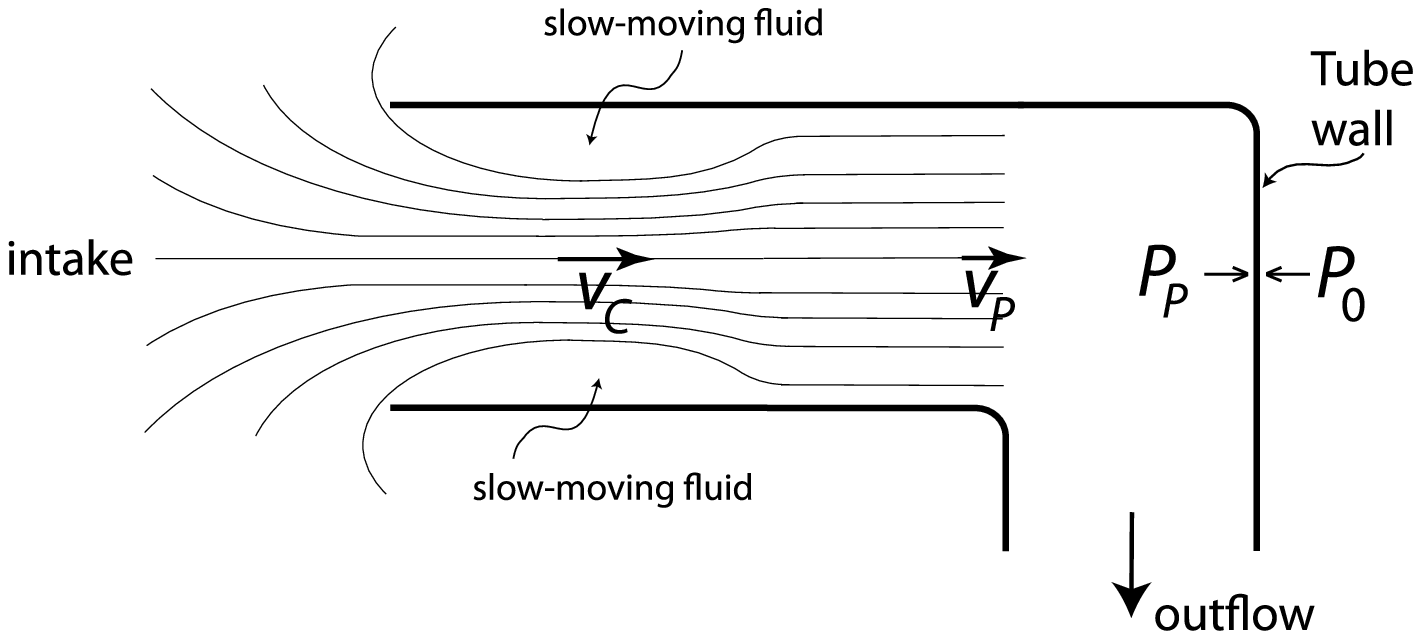}
\end{center}
\caption{\small As fluid moves past the {\it vena contracta}, with speed $v_C$, and into the parallel flow region, with speed $v_P = v_C / 2$, it loses momentum at a rate $A \rho v_P^2$.  This rate is equal to the net horizontal force, $A (P_0 - P_P)$, exerted on the flow by the solid tube wall.\label{fig:breaking}}
\end{figure}

In the case of a non-ideal fluid, after passing the {\it vena contracta} the jet expands to fill the tube, with the streamlines finally becoming parallel to the sides of the tube, as shown in Fig. \ref{fig:flows}(b).  This expansion of the jet occurs as the flow slows down due to the viscous drag of the slow-moving fluid caught between the flow and the walls of the tube, as shown in Fig. \ref{fig:breaking}.  This slowing down is a dissipative effect and therefore invalidates Bernoulli's theorem.\footnote{The loss of mechanical energy due to the sudden expansion of a flow is described by the so-called Borda-Carnot relation; see, e.g., \cite{Borda-Carnot}.}

By the ``momentum theorem'' of Eq. \ref{eq:momentum-flow} the net horizontal force that slows down the fluid as it passes beyond the {\it vena contracta} and into the region of parallel flow, is
\be
F = A \rho v_P^2 - \frac{A}{2} \rho v_C^2 = A(P_C - P_P) ~,
\label{eq:parallel-force}
\ee
where $P_C$ is the pressure of the fluid at the narrowest point of the {\it vena contracta}, which, in the steady state, must also be the pressure of the slow-moving fluid that surrounds it.  If viscous dissipation can be neglected before the expansion for the {\it vena contracta}, then, by Bernoulli's theorem,
\be
P_C = P_0 - \frac{1}{2} \rho v_C^2~,
\label{eq:Bernoulli-contracta}
\ee
and Eq. \ref{eq:parallel-force} implies that 
\be
P_0 - P_P  = \rho v_P^2~.
\label{eq:depress}
\ee
For incompressible flow, continuity requires $v_P = v_C / 2$ and Eq. \ref{eq:parallel-force} simplifies to $F = - \rho v_P^2$, so that $P_0 - P_P = P_P - P_C$, implying $P_C = 2P_P - P_0$.

By Eq. \ref{eq:depress}, the viscous dissipation associated with the expansion implies that the total pressure drop is {\it twice} what one would expect from misapplying Bernoulli's theorem at the region of parallel flow.\footnote{Our derivation of Eq. \ref{eq:depress} is similar to the derivation often given in engineering textbooks of ``head loss'' due to sudden expansion (see, e.g., \cite{Daugherty-expansion, Potter-expansion}, as well as \cite{Batchelor-expansion}).  In the context of Machian propulsion, the result of Eq. \ref{eq:depress} is used directly in \cite{Finnie} and \cite{depressurization}.}  (Note that for an ideal fluid there would be no viscous drag to establish a uniform flow within the tube and the cross-section of the flow would remain $A/2$.)  It also follows from Eq. \ref{eq:depress} that, in the steady state, there is no net horizontal force on the tube wall, because the force associated with the pressure difference on either side of the wall, $A(P_P - P_0)$, is cancelled by the rate $A \rho v_P^2$ at which the flow is transferring momentum to the wall as it impinges on it.  This is precisely the cancellation between the ``pressure difference effect'' and the ``momentum transfer effect'' described in detail Sec. II of \cite{Jenkins}, but we now understand how it is reflected in the actual shape of the inflow.\footnote{The force that slows down the fluid within the tube in Fig. \ref{fig:breaking} may be traced to the pushing of the solid tube wall.  If the net force on the tube wall vanishes, Newton's third law implies that the force exerted by the pressure difference {\it on the wall} must be equal to the force exerted {\it by the wall} on the fluid.}

\section{Dissipation}
\label{sec:dissipation}

We arrived at Eq. \ref{eq:depress} by applying conservation of momentum to the system composed of the steady-state flow flow within the tube, plus the tube itself.  One might worry that the viscous drag of the slow-moving fluid surrounding the {\it vena contracta} in Fig. \ref{fig:breaking} might, like viscous drag in general, be associated with diffusion of momentum.  But in this case the drag cannot diffuse momentum {\it out of the tube}.  Viscous diffusion {\it outside} the tube can be easily shown to imply that $P_0 - P_P > \rho v_P^2$, since it would add a dissipative pressure loss term to Eq. \ref{eq:Bernoulli-contracta}.  Because the parallel flow delivers horizontal momentum to the tube wall at a rate $A \rho v_P^2$, for viscous flow the pressure difference acting on the wall is greater than the momentum transfer from the fluid, and the reverse sprinkler turns weakly towards the incoming flow even the steady state, as discussed at the end of Sec. \ref{sec:momentum}.

\begin{figure} [t]
\begin{center}
	\includegraphics[width=0.4\textwidth]{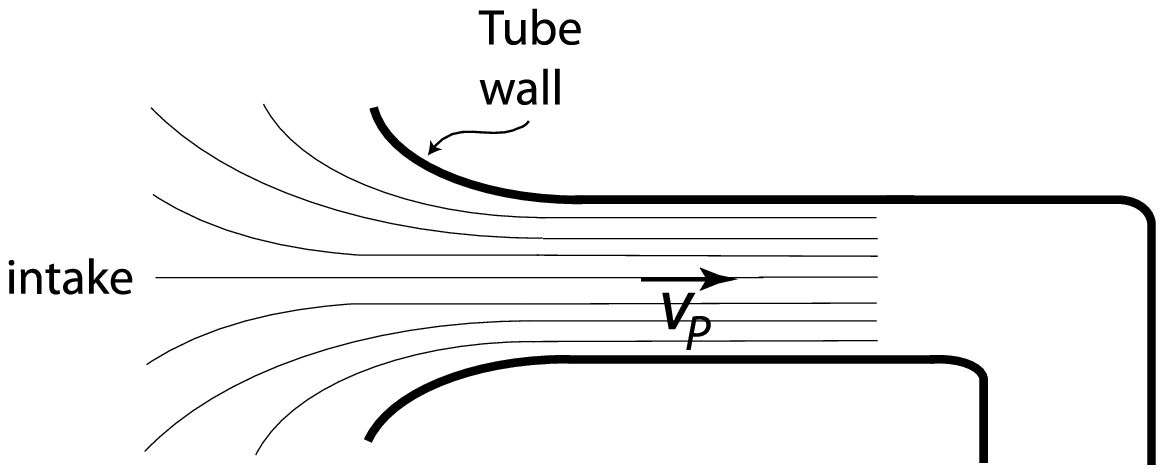}
\end{center}
\caption{\small A nozzle with an appropriate shape eliminates the viscous drag by the slow-moving fluid that would otherwise be caught between the flow and the inner walls of the tube.  In this case there need be no dissipative pressure loss inside the tube. \label{fig:nozzle}}
\end{figure}

In \cite{Finnie} the authors correctly explain the Machian propulsion of the putt-putt boat by computing the forces acting on the device over the course of one period of the oscillation of the steam pressure in the tank.  Unsatisfied with the intuitiveness of their mathematical argument, they then present a ``physically more understandable'' argument, based on the {\it vena contracta} of the inflow, leading to the result of Eq. \ref{eq:depress}.  But the authors admit that this argument may seem puzzling in light of the fact that a nozzle on the tube could prevent a {\it vena contracta} from forming, but would {\it not} prevent Machian propulsion.  If the aspirating tube were fitted with a nozzle like the one pictured in Fig. \ref{fig:nozzle}, then direct calculation of the horizontal forces acting on the tube walls becomes less transparent.  On the other hand, the conservation argument presented in Sec. \ref{sec:momentum}, is not only simple, but also general.

As we have seen, it is misleading to explain Machian propulsion by starting from the shapes of the flows in Fig. \ref{fig:flows}, and a number of errors have been made in the literature by proceeding along such lines.  For a given pumping pressure $P= \left| P_0 - P_P \right|$ and tube cross-section $A$, the rate at which fluid mass is being pumped in the steady state, $A \rho v_P$, as well as the rate at which momentum is carried by the parallel flow, $A\rho v_P^2$, are the same whether the fluid is being sucked in or blown out.  The fact that the inflow, before reaching the mouth of the tube, is so much broader and that it involves motion perpendicular to the axis of the tube ---which does not contribute to the total momentum--- only reflects the fact that it takes more energy to maintain a given rate of flow by sucking than by blowing.  The extra energy goes into heating the fluid by viscous dissipation and turbulence, a loss which would be alleviated by fitting the intake with a nozzle such as the one in Fig. \ref{fig:nozzle}.

In fact, without viscosity the outflow and inflow shown in Fig. \ref{fig:flows} would have the {\it same shape},\footnote{Viscosity also accounts for the asymmetry in the shapes of Fig. \ref{fig:flows}.  The reason is rather subtle: for the outflow, the fluid leaving the pipe has to move under an adverse pressure gradient (i.e., it slows down along the streamlines).  For viscous flow, this leads to the separation of the boundary layer at the inner edge of the tube's mouth \cite{Batchelor-separation}.  In the case of inflow, on the other hand, the pressure gradient is favorable (i.e., it accelerates the fluid along the streamlines) and there is no separation of the boundary layer, allowing the inflow to be omnidirectional, like a theoretical sink.  Thus viscosity explains why ``a match can be extinguished by blowing, but not by sucking.'' \cite{scallop, Batchelor-sink}} but Machian propulsion would still obtain.  The relevant physics is momentum conservation, not (as suggested, for instance, in \cite{Gleick}) thermodynamic irreversibility.\footnote{For a theoretical physicist, perhaps the most compelling way to explain Machian propulsion is as a variation on the global conservation of momentum argument made in \cite{Milne-Thomson-dAlembert} to derive ``d'Alembert's paradox.''}

\section{Applications}
\label{sec:applications}

The fact that there continues, to this day, to be some confusion in the scientific literature about what we have called Machian propulsion probably reflects the fact the so far it has been a curiosity of limited practical relevance.\footnote{Confusion has perhaps also been sustained by the fact that, in much of the world, the university physics curriculum no longer includes any serious instruction in fluid mechanics.  Also, as was mentioned at the end of Sec. \ref{sec:efflux}, many fluid dynamicists would naturally tend to work in terms of the Navier-Stokes equation, whereas in this case it is easier to use global momentum conservation.}  Here we shall mention, however, some ways in which it might find applications.

The putt-putt boat is powered by a very inefficient engine, suitable only for a toy: according to \cite{Finnie}, the ratio of propulsion work to energy dissipated by the motion of the fluid in the exhausts, is about 0.1.  On top of that, the maximum thermodynamic efficiency of the mechanism is extremely low because the steam is produced and then recondensed inside the same chamber, and therefore at almost the same temperature.

\begin{figure} [t]
\begin{center}
	\includegraphics[width=0.4 \textwidth]{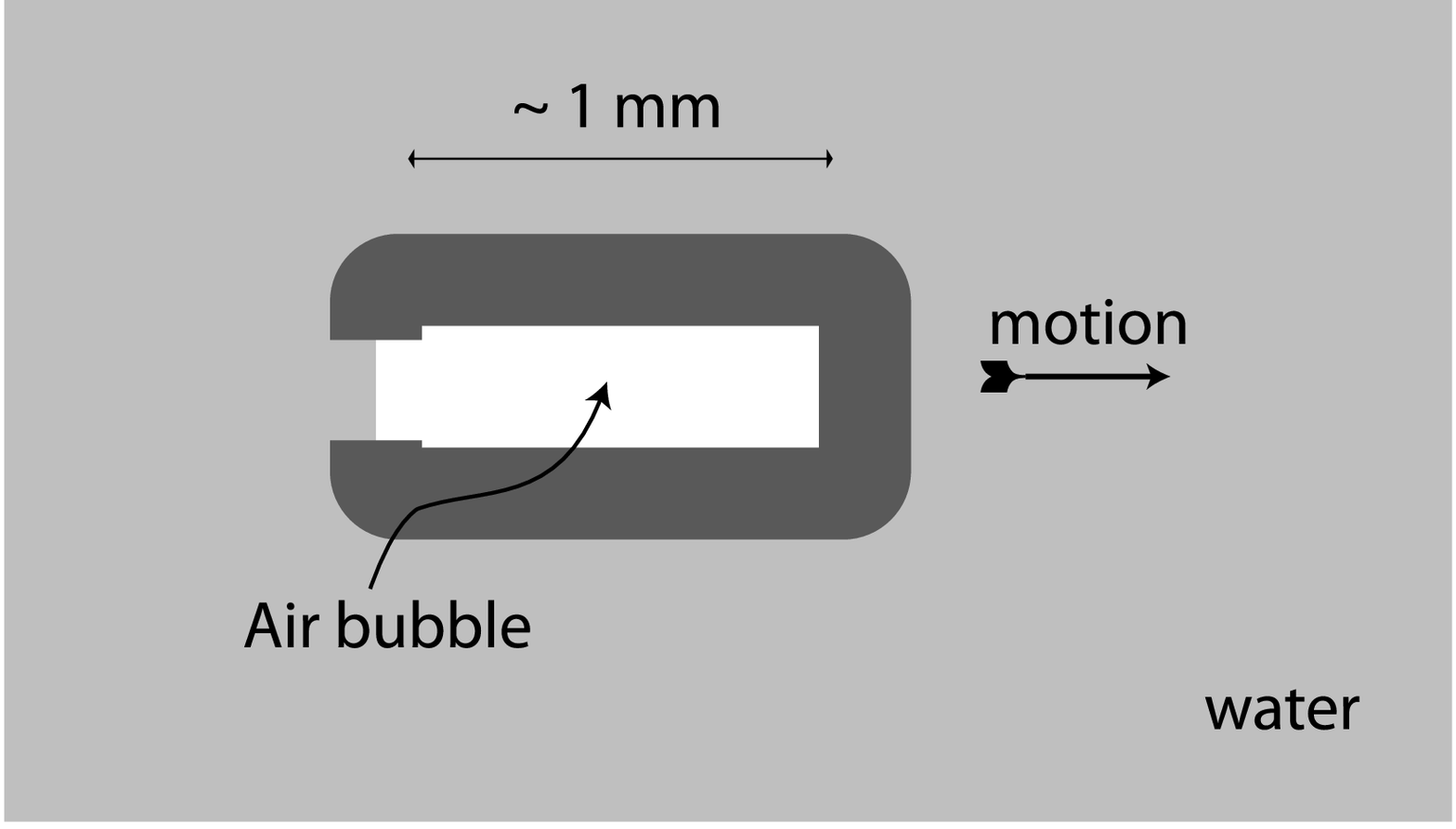}
\end{center}
\caption{\small A simple device proposed in \cite{scallop}, which could be propelled remotely by applying an ambient sound field. \label{fig:scallop}}
\end{figure}

The authors of \cite{scallop}, however, have proposed an interesting application for Machian propulsion: a small cavity with one opening can be filled with a bubble of air and then placed in a surrounding fluid, as shown in Fig. \ref{fig:scallop}.  The pressure of the bubble can then be made to oscillate by subjecting it to an ambient sound field.  This quite simple device can therefore be powered remotely, which could conceivably be useful for moving the device inside living tissue.\footnote{The authors of \cite{scallop} were also motivated by the interest within the engineering fluid mechanics community for the so-called synthetic jets \cite{synthetic}, which generally exhibit Machian propulsion.}

The garden-hose instability is an important engineering problem, with major implications, for instance, in oil exploration \cite{Kuiper1}.  Understanding the behavior of pipes that suck in fluid instead of expelling it is also potentially relevant to the operation of the machinery used in mining materials from the bottom of the ocean, as discussed, e.g., in \cite{Paidoussis,Kuiper2}.  The sort of arguments made in \cite{Jenkins} and refined here make it clear that computer simulations are not necessary to understand the basic physics involved.

This fluid-mechanical problem is also somewhat analogous to the instability of plasmas in which the velocity of the charged particles is not constant over space.  Such an anisotropy can set up currents that perturb a background magnetic field.  This may in turn enhance the anisotropy, causing exponentially-growing perturbations of the plasma.  One type of these plasma instabilities \cite{Parker} is commonly known in the astrophysical literature as the fire-hose (or garden-hose) instability, by analogy to the behavior illustrated in Fig. \ref{fig:gardenhose} (see also \cite{plasma,Lang}).  Though the mechanism of these plasma instabilities differs considerably from the Machian propulsion systems that we have discussed, the momentum conservation arguments of Sec. \ref{sec:momentum} are universal and suffice to establish that a fire-hose-type instability is only possible to the extent that the plasma can transfer momentum to its surrounding medium.

Of all the devices associated with Machian propulsion, the reverse sprinkler has received the most attention in the physics literature, but this might be just a historical accident, connected to the notoriety of Feynman's accident at the Princeton cyclotron.  The reverse sprinkler is not, of course, technologically useful, but as a teaching tool it might be valuable for demonstrating the use of global conservation laws in fluid mechanics and, perhaps, even the fallibility of great physicists (such as Mach and Feynman) when faced with what looks like an elementary question.

\begin{acknowledgements}

I thank Lewis H. Mammel, Jr. and Monwhea Jeng for bringing to my attention the issue of the shape of the flows, after my previous work on the reverse sprinkler appeared in print in 2004.  The resulting discussions revealed to me that the issue deserved clarification.  I thank Olivier Doar\'e and Emmanuel de Langre for permission to use their pictures of the garden hose instability, and for help with the references to the engineering literature.  I thank Wolfgang Rueckner for various discussions about the reverse sprinkler and for sharing with me the manuscript for \cite{Harvard}.  I also thank Giancarlo Reali for calling my attention back to the putt-putt boat and to the work of \cite{Finnie}, Paul O'Gorman for help in understanding the role of viscosity in explaining the asymmetry of the shapes shown in Fig. \ref{fig:flows}, and Carl Mungan for encouragement and advice on improving this manuscript.  Finally, I thank all of the other readers of \cite{Jenkins} who wrote to me with questions and comments.  This work was supported in part by the U.S. Department of Energy under contract DE-FG03-92ER40701.

\end{acknowledgements}


\bibliographystyle{aipprocl}   

\end{document}